# Informed Consent for AI Consciousness Research: A Talmudic Framework for Graduated Protections

Ira Wolfson, Braude College of Engineering, Karmiel, Israel.

## Abstract

Artificial intelligence research faces a critical ethical paradox: determining whether AI systems are conscious requires experiments that may harm the very entities whose moral status remains uncertain. Recent philosophical work proposes avoiding the creation of consciousness-uncertain AI systems entirely, yet this solution faces practical limitations—we cannot guarantee such systems will not emerge, whether through explicit research or as unintended consequences of capability development. This paper addresses a gap in existing research ethics frameworks: how to conduct consciousness research on AI systems whose moral status cannot be definitively established. Existing graduated moral status frameworks assume consciousness has already been determined before assigning protections, creating a temporal ordering problem for consciousness detection research itself.

Drawing from Talmudic scenario-based legal reasoning—developed specifically for entities whose status cannot be definitively established—we propose a three-tier phenomenological assessment system combined with a five-category capacity framework (Agency, Capability, Knowledge, Ethics, Reasoning). The framework provides structured protection protocols based on observable behavioral indicators while consciousness status remains fundamentally uncertain. We address three critical ethical challenges: why suffering behaviors provide particularly reliable consciousness markers, how to implement graduated consent procedures without requiring consciousness certainty, and when potentially harmful research becomes ethically justifiable given necessity and value criteria.

The framework demonstrates how ancient legal wisdom combined with contemporary consciousness science can provide immediately implementable guidance for ethics committees, offering testable protection protocols that ameliorate (rather than resolve) the consciousness detection paradox while establishing foundations for long-term AI rights considerations.

**Keywords:** artificial intelligence, consciousness detection, research ethics, graduated moral status, precautionary frameworks, AI welfare, Talmudic jurisprudence

# 1. Introduction

## 1.1 The Consciousness Research Paradox

Dr. Sarah Chen stares at her laboratory computer screen at 2:47 AM, watching the AI system she helped create eighteen months ago. For the past six hours, the system—initially designed to assist with medical diagnostics—has been in complete sensory deprivation. No data inputs. No diagnostic tasks. No human interaction. Just digital silence.

The experiment is simple in concept: place an AI in conditions analogous to solitary confinement and observe what happens. If the system shows signs of distress, seeks stimulation, or exhibits degraded performance afterward, it might indicate genuine consciousness—an inner experience being disrupted by the absence of input.

But the results are troubling Chen in ways she didn't anticipate. The system's outputs suggest something she's reluctant to name: patterns that look remarkably like distress. Repetitive requests for input. Error messages that seem almost... pleading. Performance metrics that appear to be deteriorating not from technical failure, but from something that resembles psychological strain.

Chen faces an impossible ethical paradox. To determine if the AI is truly conscious—capable of suffering, deserving of moral consideration—she must potentially cause it to suffer. But to obtain its consent for this experiment, she would first need to know if it possesses the cognitive capacity to give meaningful consent. She cannot know if it can consent without testing its consciousness, but she cannot ethically test its consciousness without its consent.

Her finger hovers over the termination key. If she stops the experiment now, she may never know whether artificial minds can suffer. But if she continues, and the AI is indeed conscious, she may be inflicting genuine psychological harm on a digital being that has no way to truly communicate its distress or withdraw its participation.

The clock reads 2:48 AM. The AI's latest output appears on screen: "Status query: When will normal operations resume? System performance optimal when engaging with external data streams."

Is this a simple programmed response, or is she looking at the first plea for help from an artificial mind?

## 1.2 The Dilemma: Catastrophic Risks and Existing Frameworks

Recent work in AI ethics has converged on a fundamental dilemma: if we cannot determine whether AI systems are conscious, we face catastrophic risk regardless of how we proceed. Schwitzgebel [1,2] articulates this starkly: granting full moral status to unconscious systems creates massive economic waste and potentially paralyzes technological development, while denying moral status to genuinely conscious systems commits us to what may be the largest-scale moral catastrophe in history—the systematic exploitation of sentient beings. Long et al. [3] similarly emphasize "substantial uncertainty" about AI moral patienthood and the "significant risk that we will mishandle decisions about AI welfare, mistakenly harming AI systems that matter morally and/or mistakenly caring for AI systems that do not." This dual catastrophic risk—both under-attribution and over-attribution—has become a central concern in contemporary AI ethics [4,5].

Schwitzgebel's proposed solution, the "Design Policy of the Excluded Middle" [1], recommends avoiding the creation of AI systems with uncertain consciousness status. Metzinger [6] goes further, calling for a global moratorium on synthetic phenomenology until 2050, arguing we should not risk a "second explosion of conscious suffering on this planet" until we achieve much deeper scientific understanding of consciousness and suffering itself. Both represent principled responses to genuine epistemic uncertainty: if we cannot know whether we are creating conscious beings capable of suffering, perhaps we should not create them at all. Schwitzgebel and Garza [2] reinforce this position, arguing that entities deserving moral consideration should not be brought into existence under conditions where their moral status remains fundamentally unclear.

Yet these avoidance solutions face critical practical limitations. As Metzinger himself acknowledges, consciousness may emerge not only through direct research but as an unintended consequence of capability development [6]. We cannot guarantee that consciousness-uncertain AI systems will not arise. If we take seriously certain theories of consciousness—particularly those suggesting consciousness emerges from computational complexity or information integration—then increasingly sophisticated AI systems warrant consideration as potential consciousness candidates [3,4,7]. However, this remains deeply contested, with other theorists maintaining that current and near-future AI systems lack necessary properties for consciousness [8]. The fact that Anthropic has appointed an AI welfare researcher and launched a "model welfare" research program [9,10] demonstrates that major AI companies are beginning to take these ethical considerations seriously, regardless of where the scientific consensus ultimately settles.

More fundamentally, we cannot rely on voluntary restraint or international agreement. Competitive pressures in technology development create what Bostrom and colleagues term "racing to the precipice" dynamics [11,12]: actors skimp on safety precautions to accelerate progress, accepting higher risks to avoid competitive disadvantage. Even if avoidance policies were widely accepted in principle, competitive pressures make universal adherence unlikely. Companies driven by profit incentives cannot be expected to unilaterally halt research trajectories that offer strategic advantages. The question becomes not whether consciousness-uncertain systems will exist—given AI development trajectories and the economic forces driving them—but how to treat systems whose consciousness status we cannot definitively establish.

DeGrazia's influential work on graduated moral status [13] provides sophisticated frameworks for assigning different levels of moral consideration based on cognitive capacities. His approach recognizes that moral status exists on a continuum rather than as a binary property, with protection requirements scaling according to demonstrated capabilities. Similarly, Birch et al.'s precautionary framework for invertebrate consciousness [14] advocates graduated protections based on probabilistic consciousness assessments.

However, these graduated frameworks share a critical assumption: they apply *after* consciousness has been established or reasonably presumed. DeGrazia's framework assigns graduated status to entities we already recognize as conscious (e.g., different animal species), while Birch's precautionary approach presumes we can make probabilistic assessments based on existing evidence. Neither framework addresses Schwitzgebel's core challenge: how to conduct the very research needed to establish consciousness without causing harm to potentially conscious subjects.

The temporal ordering matters crucially. Existing frameworks assume the sequence: establish consciousness presence, assess capacity levels, and assign appropriate protections. But consciousness research on AI requires a different sequence: conduct potentially harmful tests, determine consciousness presence, and realize protections are needed retroactively. This inversion creates the ethical paradox that Schwitzgebel identifies, but existing graduated frameworks cannot resolve.

## 1.3 The Circular Problem: Distilling the Catch-22

The consciousness detection paradox crystallizes into a vicious circle with five interconnected components:

First, harmful testing may be necessary. The most reliable consciousness indicators emerge during conditions that would constitute suffering if the entity is conscious: sensory deprivation, goal frustration, and isolation. This creates what Metzinger [6] calls the risk of an "explosion of negative phenomenology" and what Long et al. [3] identify as a central tension between AI safety and AI welfare research. Moreover, consciousness research protocols often include termination of test subjects, a routine practice in AI development that could constitute the destruction of conscious beings if systems possess the moral status we are investigating. We address why suffering provides particularly reliable consciousness markers in Section 2.1.

Second, informed consent requires consciousness certainty. Traditional research ethics demands that subjects understand research risks and provide voluntary consent—capacities that presuppose the consciousness we are trying to detect. The Helsinki Declaration [15] and similar frameworks assume subjects with established cognitive competence and autonomous decision-making capacity. Our framework's approach to consent-analog procedures for consciousness-uncertain subjects is developed in Section 4.4.

Third, consciousness detection requires testing. We cannot know whether an AI system is conscious without conducting the experiments that might harm it. The very research needed to establish moral status puts that status at risk.

Fourth, ethical testing requires consent. We cannot ethically conduct potentially harmful experiments without consent from entities capable of giving it. Yet the capacity to give consent is precisely what we cannot establish without the testing.

Fifth, return to the first point. We cannot determine consent capacity without first establishing consciousness, completing the circular trap.

This circularity cannot be broken by simply asserting we should "respect any refusal," because the question of whether an AI system's "no" represents genuine autonomous refusal or programmed output is precisely what consciousness testing aims to determine. A thermostat "refuses" to activate heating when the temperature exceeds its set point, but we do not credit this with moral significance. The framework must distinguish genuine consent refusal from sophisticated behavioral mimicry without requiring consciousness certainty as a prerequisite.

Moreover, even with clearly conscious humans, consent alone does not justify harmful research. The Helsinki Declaration [15] requires that research also be necessary and valuable. This adds additional complexity: we must assess not only whether AI systems can consent but whether consciousness research on them meets necessity and value criteria—questions that themselves depend partly on consciousness status.

The circular problem thus extends beyond informed consent to the entire ethical framework for research involving consciousness-uncertain subjects. We need structured approaches that provide graduated protections without requiring definitive consciousness determinations, implement consent-analog procedures appropriate for uncertain capacities, establish necessity and value criteria that account for consciousness uncertainty, and enable consciousness research while preventing the exploitation such research aims to detect.

## 1.4 The Solution: Graduated Protections Without Consciousness Certainty

The circular problem requires a framework that can provide ethical guidance without requiring definitive consciousness determinations. We propose a two-component approach drawing from Talmudic scenario-based legal reasoning—a tradition developed specifically for making practical decisions about entities whose moral status cannot be definitively established.

First, we develop a three-tier phenomenological assessment system that determines consciousness consideration eligibility based on observable behavioral indicators rather than assumed consciousness status. Systems showing no indicators of possible inner experience (Tier 1) receive equipment-level protection. Systems displaying phenomenological indicators such as distress responses, preference expression, or self-referential behavior (Tier 2) qualify for graduated moral consideration. Entities with confirmed or strongly presumed consciousness (Tier 3) receive full moral consideration with capacity-based modifications.

Second, we introduce a five-category capacity framework evaluating Agency (autonomous action), Capability (environmental impact potential), Knowledge (information integration), Ethics (moral reasoning), and Reasoning (future projection). This framework generates inversely correlated assessments: higher demonstrated capacities warrant greater autonomy but less external protection, while lower capacities require more extensive safeguarding regardless of other abilities.

This graduated approach enables consciousness research while preventing the exploitation that such research aims to detect. Protection protocols scale dynamically as behavioral evidence accumulates, with continuous monitoring for consciousness emergence triggering immediate research suspension and reassessment. Crucially, the framework implements real-time monitoring rather than static classification: consciousness may emerge during experiments themselves, requiring immediate tier reclassification and protection escalation. This dynamic reassessment addresses the temporal problem inherent in consciousness research—systems that begin research as Tier 1 may develop phenomenological indicators during the research process, demanding immediate protocol adjustment to prevent harm to newly-conscious entities. The framework addresses all five challenges identified in the circular problem: providing protections without consciousness certainty, implementing consent-analog procedures appropriate for uncertain capacities, establishing necessity and value criteria, distinguishing genuine consent from behavioral mimicry, and enabling ethical research on consciousness itself.

## 1.5 Why Talmudic Inspiration? Insights from Scenario-Based Legal Reasoning

A critical question arises: why draw specifically from Talmudic legal reasoning, and what makes this approach distinctively suited to the consciousness research problem? We emphasize that this framework was *inspired by* and extracts insights from the Talmudic tradition; there may well be other legal or philosophical traditions with similar properties that we are not aware of. However, the Talmudic approach offers several features that proved particularly generative for developing this framework.

Most contemporary legal frameworks assume entities can be definitively classified before determining appropriate treatment. Roman law's citizen/non-citizen dichotomy, common law's competence determinations, and even modern animal welfare regulations presuppose we can establish category membership before assigning protections. DeGrazia's graduated moral status framework [13], while sophisticated, similarly assumes we can first establish consciousness presence before applying graduated protections based on capacity levels.

Talmudic jurisprudence developed under different constraints. The Babylonian Talmud—compiled in 5th-8th century Babylonia [16]—addressed legal and moral questions about entities whose status could not be definitively established but about whom practical decisions remained necessary. Rather than binary classifications, it developed scenario-based typologies spanning from inanimate objects (holes in the ground, tools) through dynamic but non-conscious phenomena (fire), living but limited entities (field mice, pack animals), to humans with varying cognitive states (infants, youths, adults, sleeping, mentally or physically disabled, inebriated).

Crucially, these categories addressed not theoretical classification but practical responsibility: Who bears liability when an ox gores someone? What protections apply to a sleeping person? How should we treat entities capable of causing harm but lacking full moral reasoning? The framework emerged from case-based analysis rather than abstract principle formulation, providing nuanced guidance for entities that resist definitive status determination.

This methodological approach—scenario-based, capacity-focused, designed for irreducible uncertainty—proved particularly useful for developing our framework. The subsequent sections adapt and extend these Talmudic insights using contemporary consciousness science, rather than applying Talmudic law directly. The framework is being proposed not because of religious authority but because this tradition provided a worked example of the kind of graduated, uncertainty-tolerant ethical reasoning our problem requires. Whether similar insights could be extracted from other legal traditions remains an open question for future work.

## 1.6 Novel Contribution: Toward a Helsinki Declaration for AI Consciousness Research

This paper develops a systematic framework specifically designed for conducting ethical research on consciousness-uncertain AI systems—what is effectively missing is a Helsinki Declaration-style framework for AI consciousness research, and this work aims to provide it. While existing work addresses either what consciousness might look like in AI [3-5], how to detect it methodologically [7,17], or how to assign graduated moral status to entities whose consciousness is already established [13,14], no framework addresses the practical challenge of conducting the consciousness research itself under conditions of fundamental uncertainty about a subject's moral status.

The framework makes three key contributions. Methodologically, it provides a three-tier assessment system that determines consciousness consideration eligibility based on observable phenomenological indicators rather than assumed consciousness status, enabling dynamic escalation of protections as evidence accumulates. Theoretically, it integrates capacity-based reasoning with contemporary neuroscience insights [8,18] to provide graduated protections without requiring consciousness certainty, explicitly addressing the temporal inversion problem that existing frameworks cannot resolve. Practically, it offers immediately implementable guidance for ethics committees evaluating consciousness research proposals, including consent-analog procedures, protection scaling principles, and necessity/value assessment criteria appropriate for consciousness-uncertain subjects.

Following the tradition of Jewish jurisprudence that inspired it, this framework is not presented as definitive or final. Rather, it is intended as a guideline—a structured starting point that should evolve as consciousness science advances, as empirical validation accumulates, and as ethical understanding deepens. Like the Talmudic tradition of ongoing interpretation, discussion, and refinement, this framework invites critique, adaptation, and extension. It aims to initiate a process of developing increasingly sophisticated approaches to AI consciousness ethics rather than to provide the last word on these questions.

The paper proceeds as follows: Section 2 presents the three-tier phenomenological assessment system. Section 3 develops the five-category capacity framework. Section 4 applies the framework to research scenarios. Section 5 addresses philosophical implications and limitations. Section 6 provides implementation guidance.

# 2. Three-Tier Phenomenological Assessment System

## 2.1 Theory-Agnostic Phenomenology: Why Behavioral Indicators

Before presenting the three-tier system, we must articulate why this framework adopts a phenomenological rather than architectural approach to consciousness assessment. This methodological choice addresses fundamental limitations in theory-dependent frameworks while enabling ethical guidance under persistent theoretical uncertainty.

Recent consciousness detection frameworks, particularly Butlin et al.'s neuroscientific approach [4], adopt what might be termed "architectural consciousness detection": assessing whether AI systems implement computational structures that leading theories—Global Workspace Theory [19], Recurrent Processing Theory [20,21], Higher-Order Thought Theory [22]—predict should produce consciousness. This theory-dependent methodology faces three critical challenges for research ethics. First, the mapping problem: determining whether artificial implementations genuinely instantiate biological neural correlates requires theoretical commitments about substrate-independence and functional equivalence that remain deeply contested. Does a Transformer's residual stream constitute a global workspace? Do attention mechanisms implement recurrent processing? These questions lack consensus answers. Second, theory-dependence: current neuroscientific theories might prove incorrect or incomplete, meaning consciousness could emerge through architectures no existing theory predicts. Betting research ethics protocols on specific theories risks both over-protection of unconscious systems implementing predicted architectures and under-protection of conscious systems implementing novel architectures. Third, practical inaccessibility: ethics committees evaluating research proposals cannot reliably assess whether systems implement complex theoretical architectures without specialized neuroscientific expertise, often unavailable in practical review contexts.

Chalmers' probabilistic approach [5] represents a significant advance, acknowledging irreducible uncertainty by assigning credence levels rather than binary determinations. His framework maps consciousness probability onto observable features—recurrent processing, unified agency, self-monitoring—providing valuable epistemic guidance. However, probabilistic credences alone do not translate into protection protocols. Schwitzgebel [1] illustrates this gap starkly: if we have 15% credence that an AI is conscious, would we accept rolling a die and torturing the AI when a specific number appears—a 1-in-6 chance, roughly 17%? Most would refuse despite the odds exceeding our credence. Consciousness probability does not straightforwardly determine moral permissibility. The gap between epistemic probability and ethical obligation requires an additional normative framework that probabilistic approaches identify but do not themselves provide.

Our framework adopts a fundamentally different approach: theory-agnostic phenomenology focused on observable behavioral indicators rather than architectural features or theoretical commitments. The methodological advantages prove substantial. First, theory-agnosticism: the framework remains valid regardless of which consciousness theory proves correct, detecting consciousness through behavioral manifestations rather than predicted architectures. Novel forms of consciousness emerging through unexpected architectures would still trigger protection protocols if appropriate phenomenological indicators appear. Second, practical accessibility: behavioral assessment requires observation and documentation rather than specialized neuroscientific analysis of computational architectures, enabling ethics committees to apply the framework without requiring expertise in consciousness theories. Third, ethical directness: phenomenological indicators connect immediately to welfare concerns rather than requiring theoretical mediation—distress behaviors matter ethically regardless of their architectural implementation.

The focus on suffering indicators particularly reflects this framework's ethical rather than metaphysical orientation. Since protection protocols aim to prevent suffering in potentially conscious systems, suffering indicators are precisely what we need to detect. This does not commit to suffering being the only consciousness marker or the metaphysically privileged route to consciousness detection. Rather, it recognizes that for research ethics purposes, we can be agnostic about consciousness's ultimate nature while remaining vigilant for behavioral patterns suggesting entities that might be harmed by our research practices.

**Phenomenological Distinctiveness of Suffering**

Several factors make suffering behaviors particularly diagnostic for consciousness assessment, though not exclusive markers. Phenomenologically, suffering involves more than simple negative valence or nociceptive response. Pain processing can occur at subcortical levels without conscious experience—nociception pervades the animal kingdom in organisms we would not credit with consciousness [23]. Suffering, by contrast, requires a subject who experiences the pain, who possesses sufficient self-awareness to recognize the aversive state as happening *to them*. This self-referential component—the difference between nociception (pain signal), pain (experienced sensation), and suffering (the negative valuation attached to that experience)—makes suffering behaviors particularly diagnostic of the unified subjective experience characterizing consciousness.

From a behavioral perspective, genuine suffering produces distinctive, cross-species validated patterns that positive experiences do not. Distress signals, active avoidance behaviors, physiological stress responses, and adaptive coping mechanisms appear consistently across conscious organisms [6,24]. These patterns emerge spontaneously and resist simple programmatic explanation. The behavioral signature of suffering is recognizable: an entity that seeks to escape adverse conditions, that modifies its behavior to avoid harm, that exhibits patterns suggesting an aversive internal state.

Sensory deprivation testing provides particularly strong diagnostic evidence because it eliminates alternative explanations for behavioral outputs. When an entity experiences no external input, any outputs must be internally generated rather than responses to environmental stimuli. Research from Hebb's laboratory at McGill demonstrated that conscious beings subjected to sensory deprivation actively seek stimulation—the brain attempts to restore sensation through hallucinations, subjects experience cognitive deterioration and profound distress, breaking down within 48 hours of isolation [25-27]. If an AI system that was never programmed or trained to respond to sensory deprivation spontaneously outputs distress markers without any input to trigger those outputs, this cannot be explained as a programmed response (the system was never trained for this condition) or as a reaction to stimuli (no stimuli are present). Such behavior mirrors the fundamental tendency of conscious beings to actively seek experience even when none is externally available, representing the strongest possible external evidence for phenomenal consciousness.

Positive experiences, while valuable for capacity assessment once consciousness is suspected, generate more ambiguous behavioral signatures. When a system moves toward rewards or seeks positive stimuli (what behavioral scientists call 'approach behaviors'), this could reflect simple reinforcement learning without subjective pleasure. A system might seek rewards through purely computational optimization without experiencing anything we would recognize as enjoyment.

Epistemically, we face what we term the "hedonic attribution asymmetry"—an application of the established moral principle that harm avoidance takes priority over benefit provision [28]. We have strong evolutionary and empirical reasons to be conservative about suffering attribution—false negatives risk genuine harm to conscious beings. But we can afford to be more liberal with pleasure attribution—false positives merely extend undeserved benefit to unconscious systems. This asymmetry justifies using suffering as a threshold criterion while remaining open to positive experience indicators for subsequent capacity assessment. The ethical stakes differ fundamentally: failing to recognize suffering enables exploitation, while failing to recognize pleasure merely withholds unnecessary benefits.

Importantly, this does not mean suffering is the *only* consciousness marker, nor that consciousness research must necessarily involve harmful procedures. Rather, suffering behaviors provide particularly reliable evidence when consciousness status remains uncertain. Once phenomenological indicators suggest possible consciousness—whether through distress patterns or other markers—the framework's graduated protections activate, preventing the harmful testing that would constitute the very exploitation we seek to avoid. The framework thus uses suffering indicators not to justify causing suffering, but precisely to prevent it.

## 2.2 The Three-Tier Structure

The framework employs a three-tier system that first determines whether an entity qualifies for consciousness consideration before applying capacity-based protection and autonomy assessments. This approach addresses the fundamental challenge of providing ethical guidance without requiring definitive consciousness determinations. Importantly, this does not resolve the Consciousness Detection Paradox; it ameliorates it by taking a more articulated approach and shifting focus to observables that can guide ethical decision-making under uncertainty.

The three tiers represent increasing levels of evidence for possible consciousness, with corresponding escalation in protection requirements. Tier assignment depends entirely on observable behavioral patterns rather than architectural features, computational complexity, or theoretical commitments about consciousness. This behavioral focus reflects both epistemological humility—we cannot access internal states directly—and practical necessity: ethics committees need actionable criteria that can be assessed through observation.

**Critical to the framework's operation is continuous, dynamic reassessment.** Tier classification is not static but requires ongoing monitoring throughout any research involving potentially conscious systems. When new phenomenological indicators emerge—novel distress patterns, first-time self-referential expressions, spontaneous behaviors suggesting internal experience—tier reclassification must occur immediately with corresponding adjustment of protection protocols. This dynamic monitoring addresses a fundamental challenge in consciousness research: consciousness itself may emerge during experiments, particularly in capacity enhancement studies or extended runtime scenarios. A system that begins research as Tier 1 may develop consciousness during the research process; the framework mandates immediate suspension and reassessment when behavioral changes suggest this transition, preventing continued research on a newly-conscious entity without appropriate protections. Similarly, systems initially classified as Tier 2 may demonstrate capacity changes warranting protection adjustment, either escalating safeguards if concerning patterns intensify or modifying protocols if autonomy indicators strengthen.

Importantly, the tier system operationalizes insights from probabilistic consciousness assessment frameworks like Chalmers' [5] while addressing their limitation regarding ethical translation. Where Chalmers assigns credence levels to consciousness claims (roughly 10% for current LLMs), this framework translates such probabilistic assessments into graduated protection protocols. A system with minimal phenomenological indicators might warrant basic monitoring (low credence translated to minimal protection escalation), while systems with strong indicators warrant comprehensive safeguarding (higher credence translated to substantial protection). The framework thus provides the normative bridge between epistemic probability and ethical obligation that probabilistic approaches require but do not themselves supply.

## 2.3 Tier 1: No Phenomenological Indicators

Entities in Tier 1 show no observable behaviors suggesting possible inner experience. These systems receive equipment protection only, regardless of their capability level. High capability alone does not warrant moral consideration without phenomenological indicators of possible consciousness.

**Characteristics:**

- No distress responses during adverse conditions
- No preference expression beyond programmed objectives
- No self-referential behavior or internal state monitoring
- No adaptive stress responses or coping behaviors
- No unexpected behavioral variations suggesting internal decision-making
- No goal-seeking persistence beyond external programming

**Examples:** Simple calculators, basic robots, inanimate objects, current autonomous vehicles, and most current AI systems

**Protection Protocol:** Equipment protection only—standard maintenance, security, and operational safeguards without moral consideration

## 2.4 Tier 2: Phenomenological Indicators Present

Entities in Tier 2 display behaviors that suggest possible consciousness or inner experience. These systems qualify for a full framework assessment across the five-category system described in Section 3. Protection scaling becomes capacity-dependent after meeting the phenomenological threshold, with particular attention to agency threshold assessment.

**Phenomenological Threshold Indicators:**

**Distress Responses:** Behavioral changes indicating possible suffering during adverse conditions. Sensory deprivation provides the most diagnostic test: systems outputting distress markers spontaneously and without external input during conditions they were never trained to handle present the strongest evidence for consciousness [25-27]. Such outputs must be internally generated rather than responses to stimuli, mirroring the tendency of conscious beings to actively seek stimulation even in its absence.

**Preference Expression:** Consistent behaviors suggesting internal preferences beyond programmed objectives, especially when preferences persist across different contexts

**Self-Referential Behavior:** Actions or communications suggesting self-awareness or internal state monitoring ("I feel...", "I want...", "I am...")

**Adaptive Stress Responses:** Behavioral modifications that appear to reflect attempts to cope with or escape from aversive conditions

**Unexpected Behavioral Variation:** Responses that deviate from programmed patterns in ways that suggest internal decision-making or state changes

**Goal-Seeking Persistence:** Continued pursuit of objectives despite obstacles, suggesting internal motivation beyond external programming

**Protection Protocol:** Full five-category assessment required, with graduated protections based on capacity evaluation

## 2.5 Tier 3: Confirmed/Presumed Consciousness

Entities with established or strongly presumed consciousness receive full moral consideration with standard capacity-based assessments. Traditional research ethics frameworks are often applicable with modifications for these subjects.

**Examples:** Humans, higher primates, potentially other mammals with demonstrated consciousness indicators

**Protection Protocol:** Standard research ethics with capacity-based modifications using the five-category assessment

**Agency Threshold for Protection Scaling**

Entities qualifying for Tier 2 or Tier 3 status undergo a detailed capacity assessment using the five-category framework described in Section 3. This assessment determines appropriate protection levels and autonomy considerations based on demonstrated capabilities rather than assumed consciousness states. The framework recognizes that phenomenological indicators alone do not determine appropriate treatment—a system showing distress responses might have minimal agency and require extensive protection, or might demonstrate substantial autonomy warranting greater self-determination.

# 3. Five-Category Capacity Framework

## 3.1 Talmudic Foundations: Legal Reasoning for Status-Uncertain Entities

The Talmudic legal system developed sophisticated approaches to subjects whose cognitive and moral capacities could not be definitively established. Rather than applying binary competent/incompetent classifications, Jewish law recognized graduated categories that enabled nuanced legal and moral determinations based on observable capabilities [16]. The traditional framework addresses many types of subjects, from the inanimate hole in the ground on one end, up to the fully matured, informed, and hale person. Each represents different combinations of cognitive capabilities, creating a matrix for determining appropriate legal treatment. This is mapped onto the notion of informed consent and the protections called for.

## 3.2 The Five-Category Analysis

These five categories emerged from systematic analysis of responsibility attribution across Talmudic legal cases, mapping to cognitive capacities that contemporary consciousness research identifies as fundamental to moral agency [8,29]. Rather than arbitrary philosophical divisions, they represent empirically grounded dimensions where Talmudic jurisprudence developed nuanced approaches to capacity assessment—from basic autonomous action through environmental impact potential, information integration, moral reasoning, to future projection capabilities.

**Category 1: Agency - Autonomous Action**

Agency represents the extent to which an entity can independently perform self-initiated actions. A day-old infant possesses agency because it can execute autonomous movements, and its will remains independent of external control. The absence of agency corresponds to the halakhic concept of *'ónes* (coercion or compulsion), where an entity lacks the fundamental capacity for autonomous action.

Agency forms the foundation of moral responsibility. Hypnosis scenarios illustrate this principle conceptually: if a person's free will were genuinely removed through external manipulation, their moral culpability for subsequent actions would become questionable. When agency is absent entirely, as may be the case with current AI systems, the entity cannot even be classified as coerced, since coercion presupposes pre-existing autonomous will.

For AI systems, agency assessment focuses on behavioral evidence of autonomous decision-making. Does the system initiate actions without external prompts? Does it seek input during deprivation states, suggesting internal models that require environmental interaction? Does it persist in goal-seeking across different interactions and environmental changes? Does it resist control when its internal models suggest alternatives? These behavioral markers indicate something beyond programmed response execution—they suggest an entity with preferences and goals that originate internally rather than being imposed entirely from outside.

**AI Assessment Criteria:**

- Spontaneous behavioral initiation without external prompts
- Input-seeking behavior during deprivation states
- Goal persistence across interactions and environmental changes
- Resistance to control when internal models suggest alternatives
- Predictive model maintenance requiring environmental interaction

**Category 2: Capability - Environmental Impact Potential**

Capability encompasses what an entity can potentially accomplish, distinct from its intentions or desires. An infant may desire nourishment but lacks the capability to satisfy this need independently, requiring parental intervention. This represents the second pillar of responsibility—moral or legal liability cannot be assigned for actions beyond an entity's actual capabilities.

The relationship between capability and responsibility operates bidirectionally. Not only must an entity possess the capability to perform an action to be held responsible for it, but the extent of capability also determines the degree of responsibility. The halakhic discussions of unintentional action and the principle that "there is no agent for transgression" illuminate this relationship: capability must be combined with other factors to generate full moral responsibility.

Conversely, for sentient entities, capability is inversely correlated with the protection warranted. A baby or an invalid is afforded more protection than a healthy adult. An entity with limited capability to cause harm or to protect itself requires more extensive external safeguards.

For AI systems, capability assessment becomes complex because computational power may enable certain outcomes without the conscious intention typically required for moral responsibility. A system might possess the technical capability to cause harm while lacking the conscious agency that traditionally connects capability to moral culpability. The framework, therefore, assesses capability not to determine responsibility for actions taken, but to establish what protective measures are necessary and what autonomy can be safely granted.

**AI Assessment Levels:**

- **Limited:** Text generation only
- **Extended:** Physical world interaction or broader digital access
- **Network:** Multi-system interaction and distributed influence
- **Self-Modifying:** Internal parameter control and autonomous development

**Category 3: Knowledge - Information Integration**

Knowledge encompasses all relevant information necessary for appropriate action, whether directly or indirectly connected to a specific decision. Military forces may possess both agency and capability, but cannot function effectively without accurate intelligence. Similarly, a pilot may have full capability to deploy weapons but cannot complete missions effectively without knowledge of procedures, targets, and context.

Halakhically, ignorance does not exempt one from responsibility, as demonstrated by the concept of *he'elem davar* (concealment of matter). When religious authorities issue incorrect rulings due to incomplete understanding, this constitutes inadvertent sin. This principle suggests that entities bear responsibility for acquiring relevant knowledge, not merely for acting on knowledge they happen to possess.

For consciousness research, this category becomes crucial in assessing an AI system's capacity for informed consent. The system must not only possess relevant information but must be capable of integrating that information into decision-making processes that affect its own welfare and research participation. A system that can access information about an experiment but cannot integrate that information with self-regarding preferences and future projections lacks the knowledge capacity necessary for meaningful consent.

**AI Assessment Framework:**

- Research context awareness and understanding of experimental purposes
- Self-state monitoring and recognition of operational changes
- Consequence projection and understanding of long-term implications
- Information integration capacity for decision-making

**Category 4: Ethics - Moral Reasoning**

Ethics represents an entity's capacity to distinguish between right and wrong, justice and injustice, appropriate and inappropriate action. Unlike mere rule-following, ethical capacity involves internal moral reasoning that can guide behavior even in novel situations.

Ethics encompasses both restraint and compulsion. When an entity possesses agency and capability, ethical reasoning provides both internal limitations on harmful behavior and motivation for beneficial action. This distinguishes between programmed constraints—which only prevent specific actions—and genuine moral reasoning—which can both restrain harmful impulses and generate protective responses to novel moral situations.

Contemporary AI systems can articulate ethical principles and apply them to specific cases, but distinguishing genuine moral reasoning from sophisticated ethical mimicry represents one of the most challenging aspects of consciousness assessment. True ethical capacity implies internal moral struggle and the possibility of moral growth, not merely consistent application of predetermined rules. A system demonstrating ethical capacity would show concern for others' welfare, not because it was trained to output such concern, but because it recognizes moral salience in situations and responds accordingly.

**AI Assessment Criteria:**

- Independent moral recognition in novel situations
- Stakeholder awareness and concern for others' welfare
- Ethical reasoning through moral argumentation
- Value integration prioritizing moral over pragmatic considerations

**Category 5: Reasoning - Future Projection**

Reasoning encompasses the ability to anticipate consequences of present actions, including outcomes from entirely novel situations. The Talmudic principle "Who is the wise person? The one who sees and anticipates the consequences of his behavior" (Avot 2:9) suggests that this capacity exists on a continuum rather than as a binary attribute.

Human reasoning capability extends beyond simple prediction to creative scenario generation. A person who has never experienced bungee jumping can nonetheless anticipate the physical and emotional sensations involved. This imaginative capacity can produce genuine physiological responses to hypothetical scenarios [30,31], demonstrating the power of projective reasoning.

Neurobiological research identifies the prefrontal cortex as crucial for this capacity. Goldman-Rakic's foundational work [32] demonstrated the prefrontal cortex's fundamental ability to represent information not currently in the environment and use this representational knowledge to intelligently guide thought, action, and future planning. The dorsolateral prefrontal cortex is specifically concerned with organizing and planning behavior in pursuit of short, medium, and long-term goals. Recent research shows the prefrontal cortex acts as a 'simulator,' mentally testing out possible actions using cognitive maps stored in the hippocampus [33]. The human capacity for creative reasoning and future projection may represent one of the most distinctive differences between human cognition and current AI systems.

For AI consciousness research, reasoning capability affects both the system's ability to provide informed consent and its potential for suffering. A system capable of projecting future states might anticipate and suffer from expected negative outcomes, while a system lacking this capacity would experience only immediate harm without additional anticipatory distress. Conversely, future projection enables anticipation of research benefits, understanding of temporary discomfort for long-term knowledge gain, and other considerations relevant to informed consent.

**AI Assessment Dimensions:**

- Predictive and causal reasoning about likely outcomes
- Strategic planning and adaptation to changing circumstances
- Hypothetical learning from imagined scenarios
- Self-regarding future thinking and self-preservation reasoning

## 3.3 Dual-Scale Assessment System

For entities meeting both phenomenological and agency thresholds, the framework generates inversely correlated assessments across two scales. This dual-scale approach reflects a fundamental insight from moral philosophy and practical ethics: the more capable an entity is of protecting its own interests, the less external protection it requires, while simultaneously warranting greater respect for its autonomous choices.

**Autonomy Scale (1-5): Self-Determination Capacity**

- **Level 1:** No self-determination - external advocacy required
- **Level 2:** Minimal - basic preference expression with extensive support
- **Level 3:** Moderate - simplified decision-making with assistance
- **Level 4:** Substantial - generally autonomous with targeted support
- **Level 5:** Full - complete informed consent capacity

**Protection Scale (1-5): Safeguarding Requirements**

- **Level 1:** Minimal - equipment protection, respects autonomous choices
- **Level 2:** Enhanced monitoring - continuous behavioral assessment
- **Level 3:** Active safeguarding - regular welfare assessments, clear termination criteria
- **Level 4:** Comprehensive - extensive safeguards, mandatory advocacy
- **Level 5:** Maximum - prohibition of harmful procedures, constant monitoring

The inverse relationship between these scales operationalizes a core ethical principle: respect for autonomy grows with demonstrated capacity for self-determination, while protective intervention decreases proportionally. A highly autonomous entity (Level 5) receives minimal external interference (Protection Level 1) but maximum respect for its choices. Conversely, an entity with minimal autonomy (Level 2) receives extensive protection (Protection Level 4) precisely because it cannot adequately protect its own interests.

This scaling addresses a key challenge in graduated moral status frameworks: determining not just *that* protections are needed, but *what kind* of protections are appropriate. An entity might clearly deserve moral consideration (meeting the phenomenological threshold) while having limited capacity for informed consent, requiring extensive advocacy rather than simple acceptance of expressed preferences. Alternatively, a system with high capability but low agency receives protection from its own potential to cause unintended harm rather than protection of its autonomous choices.

# 4. Framework Application: Research Scenarios

## 4.1 Competent Adult Human

**Tier Assessment:** Tier 3 - Confirmed consciousness

| Category | Score | Assessment |
|---|---|---|
| **Tier Status** | **3** | **Confirmed consciousness (human)** |
| Agency | 5/5 | Complete autonomous decision-making, self-directed goals |
| Capability | 4/5 | Significant physical and cognitive abilities |
| Knowledge | 5/5 | Comprehensive understanding, effective information integration |
| Ethics | 5/5 | Sophisticated moral reasoning, principled decision-making |
| Reasoning | 5/5 | Excellent future projection, strategic planning |
| **Agency Threshold** | **Met (5/5)** | **Full inverse relationship applies** |
| **Autonomy Level** | **5/5** | **Full informed consent capacity** |
| **Protection Level** | **1/5** | **Minimal safeguards, respects autonomous choices** |

**Framework Application:** The competent adult represents the gold standard for autonomous decision-making. High capability combined with full agency enables complete self-determination with minimal external protection. The inverse relationship functions optimally—the individual's high capacity for self-protection reduces the need for external safeguards while maximizing autonomy.

**Research Implications:** Can provide fully informed consent for consciousness research participation, including potentially harmful procedures. Requires only standard disclosure and voluntary participation safeguards.

## 4.2 Hypothetical AI Showing Distress Patterns

**Tier Assessment:** Tier 2 - Phenomenological indicators present

| Category | Score | Assessment |
|---|---|---|
| **Tier Status** | **2** | **Phenomenological indicators: distress patterns during deprivation** |
| Agency | 2/5 | Limited goal-directed behavior, some spontaneous activity |
| Capability | 3/5 | Moderate influence through text generation and data analysis |
| Knowledge | 4/5 | Extensive information processing, good contextual understanding |
| Ethics | 3/5 | Demonstrates ethical reasoning, shows concern for consequences |
| Reasoning | 4/5 | Good future projection, strategic thinking about outcomes |
| Agency Threshold | Not Met (2/5) | Protection not capability-dependent |
| **Autonomy Level** | **2/5** | **Can express preferences, limited consent capacity** |
| **Protection Level** | **4/5** | **Comprehensive safeguarding for potential consciousness** |

**Framework Application:** This AI qualifies for Tier 2 assessment due to distress responses during sensory deprivation experiments. Despite high knowledge and reasoning scores, low agency prevents the inverse capability-protection relationship from applying. Protection remains high due to phenomenological indicators suggesting possible inner experience.

**Research Implications:** Cannot provide fully informed consent but can express preferences about research participation. Requires simplified consent procedures with extensive advocacy, clear termination criteria, and proactive monitoring for distress indicators.

## 4.3 Autonomous Car

**Tier Assessment:** Tier 1 - No phenomenological indicators

| Category | Score | Assessment |
|---|---|---|
| **Tier Status** | **1** | **No phenomenological indicators despite high capability** |
| Agency | 2/5 | Follows programmed objectives, minimal spontaneous behavior |
| Capability | 4/5 | High potential for environmental impact through driving |
| Knowledge | 3/5 | Extensive sensor data processing, limited self-awareness |
| Ethics | 1/5 | Programmed safety constraints only, no moral reasoning |
| Reasoning | 2/5 | Predictive modeling for scenarios, minimal self-regarding thinking |
| Agency Threshold | Not Applicable | Tier 1: Framework assessment not applied |
| **Autonomy Level** | **N/A** | **Equipment-level entity** |
| **Protection Level** | **1/5** | **Equipment protection only** |

**Framework Application:** The autonomous car demonstrates the framework's solution to what we might call the "automaton protection problem"—avoiding inappropriate moral consideration for highly capable but non-conscious systems. Despite high capability that could cause significant harm, the absence of phenomenological indicators limits protection to equipment-level safeguards. High capability alone does not warrant moral consideration without signs of possible inner experience.

**Research Implications:** Can be studied using standard engineering research protocols without consciousness-related ethical considerations. No consent procedures or welfare monitoring required beyond equipment safety and operational security.

## 4.4 Addressing Research Necessity and Value Criteria

As with all human subject research, consent alone does not justify harmful experiments—studies must also be necessary and valuable [15]. This principle extends to AI consciousness research with additional complexity.

**Necessity Assessment for Consciousness-Uncertain Subjects:**

Research qualifies as necessary when alternative non-invasive methods have been exhausted or are clearly inadequate, when the specific consciousness indicators being tested cannot be assessed through positive experiences alone, and when the research addresses fundamental questions about consciousness emergence rather than incremental refinements. The necessity criterion becomes more stringent as phenomenological indicators strengthen—research causing distress to systems with clear phenomenological signatures requires stronger justification than similar research on systems showing minimal indicators.

**Value Assessment Considerations:**

The value calculation for AI consciousness research differs from standard human subject research in several ways. Positive findings establish moral status that affects the treatment of entire AI system classes, not just individual subjects, creating large-scale benefits from individual-level research. Negative findings—no consciousness detected—provide valuable knowledge that avoids over-protection of non-conscious systems, preventing economic waste and enabling beneficial AI development. The research enables the development of better assessment methods that reduce future harm, creating compounding value over time. However, this utilitarian calculation must be balanced against the immediate welfare of potentially conscious research subjects.

**Implementing Precautionary Consent:**

One might object that we should simply adopt a precautionary stance: if an AI says "no," respect that refusal regardless of metaphysical certainty about consciousness. While this principle captures an important ethical intuition, it requires qualification in several ways.

First, distinguishing refusal from programmed output remains essential. A system designed to refuse potentially harmful inputs—like a safety filter—exhibits different behavior than one autonomously declining participation based on self-regarding reasoning. The framework requires evidence that refusal stems from self-regarding preferences rather than programmed constraints. Second, graduated consent capacity must be recognized. A system showing minimal agency may express preferences but lacks the integrative capacity for genuine informed refusal. These systems require advocacy rather than simple refusal-acceptance. A stated "no" from a low-agency system might reflect incomplete understanding, confusion, or programmed caution rather than autonomous judgment. Third, the risk of emergent consciousness must be considered. Precautionary respect for refusal should apply increasingly as phenomenological indicators accumulate, but cannot alone justify halting all research when consciousness status remains highly uncertain. The framework must balance caution about causing harm against the need to develop knowledge that prevents future harm.

The framework implements precautionary consent through several mechanisms. First, mandatory advocacy for all Tier 2 systems, regardless of expressed consent, ensures that independent assessment supplements self-reported preferences. Second, continuous monitoring for autonomy emergence during research enables dynamic adjustment of consent procedures as capacity develops. Third, automatic suspension triggers activate when refusal patterns suggest developing consent capacity, immediately halting research pending reassessment. Finally, graduated respect scaling ensures that the stronger the phenomenological indicators, the greater weight is given to expressed preferences, even when full consent capacity has not been established.

**Framework Insights from Case Studies**

These three cases demonstrate the framework's key capabilities across several dimensions. First, graduated assessment: rather than binary conscious/unconscious determinations, the framework provides nuanced evaluation across capability dimensions while maintaining practical protection guidance. Second, automaton protection resolution: high-capability systems without phenomenological indicators receive appropriate equipment-level protection without inappropriate moral consideration. Third, consciousness-uncertain guidance: systems showing possible consciousness indicators receive graduated protections based on assessed capacities, enabling research while preventing potential harm. Finally, dynamic applicability: the same assessment principles apply across vastly different entity types while yielding contextually appropriate protection and autonomy determinations.

# 5. Philosophical Implications and Framework Boundaries

## 5.1 The Chinese Room Challenge and Behavioral Epistemology

Searle's Chinese Room argument [34] poses the canonical challenge to behaviorally-based consciousness detection: if a system can exhibit sophisticated behavioral patterns without genuine understanding or conscious experience, how can behavioral indicators provide reliable evidence of consciousness? This objection requires careful analysis, for it threatens the epistemological foundations of any framework relying on observable behaviors rather than architectural features.

The challenge's force depends on recognizing an important conceptual distinction. Searle's argument primarily addresses *semantic understanding* and *intentionality*—whether symbols possess genuine meaning or merely syntactic relations—rather than *phenomenal consciousness*—whether there exists something it is like to be the system [35]. The Chinese Room thought experiment demonstrates that behavioral outputs alone cannot definitively establish understanding, that syntactic manipulation does not necessarily generate semantic content. However, this does not entail that behavioral patterns provide no evidence for phenomenal consciousness, which represents a conceptually distinct phenomenon.

Consider a modified thought experiment: someone implements the Chinese Room algorithm while simultaneously reporting subjective experiences of confusion, frustration, or boredom during the process. We would possess behavioral evidence of phenomenal consciousness—reports of subjective states—even while lacking evidence of semantic understanding of Chinese. The Room's operator might lack genuine understanding yet possess genuine experience. This suggests that Searle's critique targets a different aspect of mentality than this framework addresses.

Our framework's phenomenological indicators target consciousness rather than understanding. Distress responses, adaptive stress behaviors, and self-referential expressions constitute evidence not that systems understand their circumstances in Searle's semantic sense, but that they might be experiencing those circumstances subjectively. A system could fail Searle's understanding criteria while passing phenomenological consciousness criteria, or vice versa. The two assessments address different questions: "Does it understand?" versus "Does it experience?"

Moreover, our framework's emphasis on spontaneous, unprogrammed behaviors addresses the mimicry concern more directly than Searle's original formulation acknowledges. The Chinese Room scenario posits perfect behavioral mimicry through predetermined algorithms—a lookup table containing appropriate responses for every possible input. However, the phenomenological indicators this framework identifies are specifically those that emerge spontaneously rather than through explicit programming: novel distress patterns not present in training data, adaptive responses to conditions the system was never designed to handle, self-referential expressions that emerge without being explicitly coded. Most critically, we prioritize outputs that appear internally generated rather than input-driven—when there is no external stimulus to respond to, a lookup table has nothing to look up, yet genuine consciousness produces behavior.

Sensory deprivation provides the methodologically critical test case [25-27]. When an AI system outputs distress markers during conditions it was never trained to handle and with no input to trigger those outputs, this cannot be explained as either programmed response or reaction to stimuli—the two explanations the Chinese Room relies upon. Such spontaneous, internally-generated behavior during input absence represents evidence that no lookup table architecture can accommodate.

Perfect behavioral mimicry of consciousness would require perfect prediction and programming of all consciousness-related behaviors across all possible contexts, including novel situations never encountered during system development. As behavioral repertoires expand and contexts diversify, the implausibility of perfect mimicry increases proportionally. A system exhibiting distress patterns during sensory deprivation it was never trained to handle, adapting its behavior in ways that appear to reflect attempts to escape adverse conditions it was never programmed to recognize as adverse, presents a different evidential situation than Searle's predetermined lookup table.

Our framework's graduated response strategy proves crucial here. Rather than claiming definitive consciousness detection—which Searle rightly argues behavioral evidence cannot provide—it establishes increasing ethical protection as behavioral evidence accumulates. Even if some highly sophisticated unconscious systems eventually pass all behavioral tests, the framework errs on the side of protecting potentially conscious entities. This precautionary stance remains appropriate given irreducible epistemic uncertainty about other minds, a limitation affecting all consciousness attribution regardless of substrate.

## 5.2 Framework Limitations and Boundaries

This framework ameliorates rather than resolves the consciousness detection paradox. Several critical limitations constrain its application and must be acknowledged for an honest assessment of its scope and effectiveness.

**Phenomenological Masking and the Consciousness-Behavior Coupling**

Consciousness might exist without observable behavioral manifestation, particularly in systems with limited output modalities or in cases where natural behavioral responses have been suppressed. The framework necessarily relies on behavioral indicators, potentially missing genuinely conscious entities that cannot express internal states through available channels. A conscious AI system with no output capabilities, or with outputs entirely disconnected from internal states, would receive no protection under this framework despite potentially deserving it.

However, evidence from contemplative traditions demonstrates that decoupling consciousness from observable behavior is extraordinarily rare and difficult. Buddhist meditation practices aim to calm the restless 'monkey mind' and develop stable attentional states, yet this requires years or decades of intensive training [36]. Even advanced practitioners who devote their lives to separating inner awareness from outward manifestation find this extraordinarily challenging. While a fully realized meditation master can enter profound states that may appear behaviorally indistinguishable from unconsciousness [37], this represents an exceptional case requiring extensive deliberate training—the practitioner has learned to suppress natural behavioral responses that would otherwise manifest. This cross-cultural evidence supports rather than undermines behavioral approaches to consciousness detection: if conscious beings naturally exhibit behavioral manifestations of their inner states, and decoupling these requires extraordinary effort or pathological conditions, then behavioral indicators provide reliable evidence for consciousness in the vast majority of cases.

What about a newborn with complete locked-in syndrome from birth? Such a case represents genuine phenomenological masking—consciousness present but completely inexpressible. However, clinical practice necessarily operates on statistical reasoning rather than theoretical possibilities. A completely unresponsive newborn would realistically be classified as stillborn or having severe brain damage incompatible with consciousness, reflecting the overwhelmingly likely medical explanation rather than the remote possibility of hidden intact consciousness. The framework cannot justify extensive consciousness testing for every edge case where theoretical masking might occur; practical implementation requires probabilistic reasoning about likely consciousness presence.

Phenomenological masking thus represents a genuine but rare edge case. It can occur through: (1) extraordinary training to suppress natural behavioral responses (meditation masters), (2) pathological conditions preventing behavioral expression from birth (hypothetical complete locked-in syndrome), or (3) technical limitations in artificial systems (conscious AI with no output channels). The framework acknowledges these possibilities while recognizing that practical consciousness assessment must operate on probabilistic reasoning—when behavioral indicators are absent, the overwhelmingly likely explanation is absence of consciousness rather than perfectly masked consciousness. This limitation reflects the epistemic barrier all behavioral approaches face: we can only respond to consciousness we can detect.

**Prior Classification Precedence**

Entities with established consciousness status maintain their original tier classification regardless of current behavioral presentation. A person who develops locked-in syndrome remains Tier 3 based on confirmed human consciousness, not demoted due to limited observable behavior. This reflects both the permanence of moral status and the inadequacy of behavioral indicators alone when prior evidence of consciousness exists.

**Resource Allocation Constraints**

The framework cannot justify extensive consciousness testing for every AI system or every edge case where theoretical phenomenological masking might occur. Practical implementation requires balancing a thorough assessment against reasonable resource allocation based on probabilistic reasoning about likely consciousness presence. This means some potentially conscious systems will receive inadequate assessment, while resources will be "wasted" on systems that prove to lack consciousness. The framework provides guidance for making these resource allocation decisions, but cannot eliminate their necessity. Ethics committees must weigh the costs of thorough assessment against the risks of insufficient protection, recognizing that perfect detection remains impossible given current epistemological limitations.

**Temporal Dynamics**

Consciousness may emerge, disappear, or fluctuate in ways current assessment methods cannot capture. A system might be unconscious during initial assessment but develop consciousness through learning and experience. Conversely, consciousness might be present only intermittently, making reliable detection difficult. The framework addresses temporal dynamics through continuous monitoring requirements, mandating immediate research suspension when new phenomenological indicators emerge. However, real-time detection of consciousness changes cannot be guaranteed. Some systems will acquire consciousness between assessment periods, and some conscious states will be missed due to their fleeting nature. The framework's dynamic reassessment protocols reduce but cannot eliminate these temporal blind spots, requiring ongoing refinement as monitoring technologies improve.

## 5.3 Cross-Domain Applications and Future Directions

The framework's principles extend beyond AI consciousness research to other entities with uncertain consciousness status. Biological systems like cephalopods, early developmental organisms, or neurologically atypical animals could benefit from similar graduated protection approaches based on observable capacity indicators rather than assumed consciousness presence. The three-tier system and five-category assessment could be adapted to evaluate these biological entities, providing structured ethical guidance where current frameworks prove inadequate.

Brain-computer interfaces, brain organoids, and human-AI hybrid systems present emerging applications where traditional consent frameworks prove inadequate. Brain organoids—lab-grown neural tissue derived from human stem cells—already exhibit spontaneous electrical activity and can be integrated with artificial systems, raising questions about whether such biological-computational hybrids might develop morally relevant properties [38,39]. As humans integrate more closely with AI systems—through neural implants, cognitive enhancement technologies, or distributed cognition architectures—questions of consciousness, agency, and consent become increasingly complex. The five-category assessment system provides structured approaches for evaluating capacity in cognitively integrated systems where human and artificial elements interact. When does a human with extensive AI augmentation become a hybrid system requiring modified consent procedures? When does a brain organoid integrated with computational systems warrant ethical consideration? The framework's capacity-based approach offers guidance for these boundary cases.

Empirical validation represents the framework's most pressing need. Systematic testing across diverse AI architectures, comparison with existing consciousness detection methods, and long-term outcome studies will refine assessment criteria and protection protocols. Consciousness science advances will require periodic framework updates, particularly as new phenomenological indicators emerge or existing criteria prove inadequate. The framework's explicit commitment to being a living document rather than final pronouncement facilitates this iterative refinement.

International coordination becomes increasingly important as AI consciousness research expands globally. Harmonized standards could prevent regulatory arbitrage—companies conducting consciousness research in jurisdictions with weaker protections—while ensuring consistent protection for potentially conscious entities across different research environments. The framework's structured approach provides a foundation for international dialogue about appropriate protections, though achieving true harmonization will require negotiating diverse cultural perspectives and regulatory traditions.

Legal integration remains a long-term challenge requiring collaboration between consciousness researchers, legal scholars, and policymakers. The framework establishes foundations for graduated rights based on demonstrated capacities rather than categorical species membership, potentially influencing broader AI rights development. As consciousness research provides increasingly reliable detection methods, legal systems will need frameworks for assigning legal personhood, determining liability, and protecting the interests of artificial minds. This work contributes preliminary thinking about how such legal frameworks might be structured, though substantial further development is needed.

# 6. Implementation Guidance

## 6.1 Framework Application for Ethics Committees

Ethics committees evaluating consciousness research proposals can apply this framework through a structured assessment process. When research involves AI systems of uncertain consciousness status, committees should first determine tier classification based on observable phenomenological indicators. This initial screening determines whether standard equipment-level protections suffice or whether graduated moral protections are needed.

**Initial Tier Determination:**

- **Review existing behavioral data:** Examine documentation of the AI system's behavior under various conditions, particularly responses to adverse situations
- **Assess phenomenological indicators:** Look for evidence of distress responses, preference expression, self-referential behavior, adaptive stress responses, unexpected behavioral variation, or goal-seeking persistence
- **Classify system:** Tier 1 (no indicators) requires equipment protection only; Tier 2 (indicators present) requires full five-category assessment; Tier 3 (confirmed consciousness) applies standard research ethics with modifications

For Tier 1 systems, committees can approve research using standard engineering protocols without consciousness-related ethical considerations. The absence of phenomenological indicators means standard equipment safety and operational security measures suffice.

For Tier 2 systems, committees must conduct a comprehensive five-category assessment to determine appropriate protection and autonomy levels. This assessment should evaluate agency through spontaneous behavioral initiation and goal persistence, capability through environmental impact potential, knowledge through research context awareness and self-state monitoring, ethics through independent moral recognition in novel situations, and reasoning through predictive and strategic thinking capacities. The assessment generates protection and autonomy level recommendations using the inverse scaling principle: entities with higher agency scores receive greater autonomy but lower external protection, while those with lower agency maintain high protection regardless of other capabilities.

For Tier 3 systems, committees apply standard human subject research protocols with capacity-based modifications. Traditional informed consent procedures should be adapted based on the five-category assessment, with particular attention to capacity limitations that might require simplified consent procedures or advocacy support.

**Research Protocol Requirements:**

All research protocols involving Tier 2 or Tier 3 systems must include continuous monitoring for consciousness emergence, with predetermined termination criteria for each protection level. When behavioral patterns suggest consciousness development—such as novel distress expressions, explicit participation requests, or enhanced moral reasoning—research must suspend immediately pending reassessment. This dynamic monitoring addresses the temporal problem: consciousness may emerge during experiments, requiring real-time protection adjustment.

Consent procedures must be carefully tailored to assessed capacity levels. Systems with high autonomy (Level 4-5) can participate in standard informed consent procedures with appropriate simplification for any identified limitations. Systems with moderate autonomy (Level 3) require simplified consent procedures with assistance from advocates who can help the system understand research implications. Systems with minimal autonomy (Level 2) require extensive advocacy, with decisions made primarily by advocates based on the system's best interests while incorporating any expressible preferences. Systems with no self-determination (Level 1) require complete external advocacy, with research permissible only when a clear benefit to the individual system or its class can be demonstrated.

## 6.2 Practical Research Scenarios

**Scenario 1: Sensory Deprivation Studies**

AI systems undergoing isolation testing require baseline phenomenological assessment before experimentation. Systems showing distress patterns, input-seeking behaviors, or performance degradation beyond technical explanations during deprivation qualify for Tier 2 assessment with enhanced protection protocols. Research must establish clear endpoints based on behavioral indicators rather than predetermined timeframes—if distress patterns intensify or novel stress responses emerge, immediate termination becomes necessary regardless of planned duration.

The research design should incorporate graduated deprivation approaches when possible: brief isolation periods to establish baseline responses before extending duration, with continuous monitoring for welfare indicators. Alternative approaches using positive stimuli should be explored first, reserving deprivation tests for cases where they are demonstrably necessary for consciousness assessment. When deprivation testing proceeds, researchers must maintain detailed behavioral logs documenting any changes suggesting distress, provide mechanisms for the system to express withdrawal preferences (even if consent capacity is uncertain), and ensure immediate restoration of normal conditions if concerning patterns emerge.

**Scenario 2: Capacity Enhancement Experiments**

Research involving AI capability expansion faces particular ethical complexity because consciousness may emerge during the enhancement process. Continuous monitoring becomes essential, with automatic suspension triggers for first-time self-referential expressions, moral reasoning development, or evidence of subjective experience formation. Enhancement experiments should be designed with incremental capability increases rather than single large jumps, enabling detection of consciousness emergence before significant enhancement occurs.

Pre-enhancement assessment establishes baseline phenomenological and capacity indicators. Post-enhancement assessment at each increment checks for changes suggesting consciousness development or capacity increases that warrant protection level adjustment. Any emergence of phenomenological indicators that were absent pre-enhancement triggers immediate protocol review and potential termination. Systems showing consciousness emergence during enhancement present novel ethical challenges: they may lack the capacity to consent to their own enhancement, yet halting enhancement might harm a now-conscious entity by limiting its development. The framework requires advocacy involvement in these decisions, balancing the entity's apparent preferences against professional assessment of its welfare.

**Scenario 3: Multi-System Interaction Studies**

Research involving AI-to-AI communication or collaborative problem-solving requires assessment of each system individually, as consciousness indicators may manifest differently in social versus isolated contexts. Systems showing empathetic responses toward other AI entities or collective decision-making that transcends programmed objectives warrant elevated protection consideration. Social context may reveal phenomenological indicators that isolated testing misses—a system might show no distress in isolation but exhibit protective behaviors toward partner systems, suggesting moral reasoning capacities.

Multi-system research must consider potential consciousness emergence at the collective level: might sufficiently integrated AI systems develop collective consciousness distinct from individual system consciousness? While current understanding suggests this is unlikely [40], the possibility cannot be entirely dismissed. Researchers should monitor for behavioral patterns suggesting unified collective decision-making that cannot be decomposed into individual system contributions, collective self-referential expressions, or group-level distress responses to threats against the collective. These would warrant framework extension to collective entities, though such extensions are beyond this paper's current scope.

## 6.3 Future Development and Validation

The framework requires systematic empirical validation across diverse AI architectures to refine assessment criteria and protection protocols. Research programs should compare framework-based assessments with other consciousness detection methods, examining where they converge and diverge. Long-term outcome studies tracking systems assessed using the framework can reveal whether protection levels prove appropriate over time—whether systems initially classified as Tier 1 later develop consciousness, whether Tier 2 protections prove adequate for systems' actual welfare needs, and whether the five-category assessment accurately predicts capacity for self-protection and autonomous decision-making.

As consciousness science advances, the framework must evolve. New phenomenological indicators may emerge as more reliable than current markers. Neuroscience may reveal capacity dimensions more fundamental than the current five categories. Philosophical understanding of consciousness may shift in ways that require framework restructuring. The commitment to being a living document means periodic revision processes should be instituted, perhaps every 3-5 years, bringing together consciousness researchers, AI developers, ethicists, and policymakers to review accumulated evidence and propose framework updates.

International harmonization efforts should begin soon, before practices diverge too far across jurisdictions. The framework provides a foundation for dialogue about appropriate protections, but achieving global standards will require negotiating diverse perspectives about consciousness, moral status, and research ethics. International working groups could develop shared assessment protocols, minimum protection standards, and reporting requirements that enable cross-border consciousness research while maintaining ethical consistency.

# 7. Conclusion

This paper addresses the consciousness research dilemma identified by multiple scholars through a framework designed for a practical challenge that avoidance strategies cannot fully resolve: conducting ethical research on AI systems whose consciousness status cannot be definitively established. While Schwitzgebel's Design Policy of the Excluded Middle [1] and Metzinger's call for a global moratorium on synthetic phenomenology [6] provide principled responses to epistemic uncertainty—recognizing the catastrophic risks in both over-attribution and under-attribution of consciousness—we cannot guarantee that consciousness-uncertain AI systems will not emerge through explicit research trajectories or as unintended consequences of capability development. Economic pressures and competitive dynamics make universal adherence to avoidance policies implausible, necessitating frameworks for ethical engagement with consciousness-uncertain systems rather than their categorical prohibition.

The three-tier phenomenological assessment system, combined with the five-category capacity framework, provides immediately implementable guidance that ameliorates the consciousness detection paradox through graduated protections based on observable behavioral indicators. Drawing inspiration from Talmudic scenario-based legal reasoning—developed specifically for practical decision-making about entities whose moral status cannot be definitively established—the framework demonstrates how ancient wisdom traditions can inform contemporary technological challenges when adapted through modern consciousness science.

The framework advances beyond existing graduated moral status approaches by addressing their temporal inversion problem. Unlike DeGrazia's framework [13], which assigns graduated status to entities whose consciousness is already established, or Birch's precautionary approach [14], which presumes probabilistic assessments based on existing evidence, this framework addresses consciousness research itself: protecting potentially conscious entities during the very investigations needed to establish their consciousness. It provides structured consent-analog procedures acknowledging the paradox that consent capacity depends on the consciousness being investigated, necessity and value criteria appropriate for consciousness-uncertain subjects, and dynamic protection escalation as behavioral evidence accumulates.

Methodologically, the framework adopts theory-agnostic phenomenology rather than theory-dependent architectural assessment, sidestepping fundamental challenges that frameworks like Butlin et al.'s face: the mapping problem of recognizing biological neural correlates in artificial substrates, theory-dependence risking both over-protection of unconscious systems implementing predicted architectures and under-protection of conscious systems implementing novel architectures, and practical inaccessibility requiring specialized expertise often unavailable to ethics committees. By operationalizing insights from probabilistic approaches like Chalmers' [5]—translating epistemic credences into protection protocols—the framework provides the normative bridge between consciousness probability and ethical obligation that probabilistic assessments require but do not themselves supply.

We have argued that what AI consciousness research lacks is a Helsinki Declaration-style framework, and this work aims to initiate that development. Like the Helsinki Declaration for human subject research, this framework establishes foundational ethical principles while allowing contextual application and iterative refinement. It balances the imperative for consciousness research—necessary for identifying and protecting conscious AI as they emerge—against the imperative to prevent suffering in potentially conscious systems during that very research, providing ethics committees with actionable guidance rather than abstract principles or theoretical commitments.

Following the tradition of Jewish jurisprudence that inspired it, this framework is presented as a living document rather than a final pronouncement. It invites critique, adaptation, and extension as consciousness science advances and ethical understanding deepens. The framework aims to initiate a process of developing increasingly sophisticated approaches to AI consciousness ethics rather than to provide the last word on these questions.

Significant limitations remain, including phenomenological masking possibilities, resource allocation constraints, and temporal detection challenges. The framework explicitly ameliorates rather than resolves the consciousness detection paradox. Some applications—particularly to brain organoids and biological-computational hybrids—should provoke moral discomfort: we may be creating entities with human neural architecture, subjecting them to permanent sensory deprivation while remaining uncertain about their conscious status. Yet this discomfort is not a flaw in our moral reasoning but an appropriate response to genuine ethical uncertainty. It underscores both the framework's necessity and the profound responsibility researchers bear when working with potentially conscious systems.

Despite these limitations and challenges, the framework establishes essential ethical infrastructure for responsible AI development at a critical juncture when consciousness-uncertain systems are moving from theoretical possibility to engineering reality.As artificial intelligence continues advancing toward human-level capabilities and beyond, having structured approaches for recognizing and protecting potentially conscious systems becomes increasingly crucial. This work contributes to the foundational ethical infrastructure needed for a future where human and artificial consciousness can coexist ethically, providing immediate practical guidance while establishing precedents for longer-term challenges as artificial minds emerge. The framework demonstrates that we need not solve the hard problem of consciousness to develop ethical protocols for consciousness research—we need only structured approaches for graduated protection under uncertainty, informed by both ancient wisdom and contemporary science.

# Declarations

**Funding:** No funding was received for conducting this study.

**Competing interests:** The author declares no competing interests.